\documentclass[useAMS,usenatbib,usegraphicx]{mn2e}
\newcommand{\pdt}[1]{{{\partial #1}\over {\partial t}}}
\newcommand{\Mp}{M_p}

\newcommand{\kape}{\kappa_{e}}
\newcommand{\fup}{F_{\uparrow,b}}
\newcommand{\fdown}{F_{\downarrow,b}}

\title[Radiative Hydrodynamical Simulations of HD189733b]{Three
  Dimensional Radiative Hydrodynamical Simulations of the Highly
  Irradiated Short Period Exoplanet HD189733b} \author[Ian Dobbs-Dixon
  and Eric Agol]{Ian Dobbs-Dixon$^{1,2}$\thanks{
    iandd@astro.washington.edu (IDD); agol@astro.washington.edu (EA)}
  and Eric Agol$^{1}$\\$^{1}$Department of Astronomy, Box 351580,
  University of Washington, Seattle, WA 9819\\$^{2}$$^2$NASA
  Astrobiology Institute}

\begin{document}

\date{Submitted 2012 November 5}

\pagerange{\pageref{firstpage}--\pageref{lastpage}} \pubyear{2012}

\maketitle

\label{firstpage}

\begin{abstract}
We present detailed three-dimensional radiative-hydrodynamical models
of the well known irradiated exoplanet HD189733b. Our model solves the
fully compressible Navier-Stokes equations coupled to
wavelength-dependent radiative transfer throughout the entire
planetary envelope. We provide detailed comparisons between the
extensive observations of this system and predictions calculated
directly from the numerical models. The atmospheric dynamics is
characterized by supersonic winds that fairly efficiently advect
energy from the dayside to the nightside. Super-rotating equatorial
jets form for a wide range of pressures from $10^{-5}$ to $\sim 10$
bars while counter rotating jets form at higher latitudes. Calculated
transit spectrum agree well with the data from the infrared to the UV
including the strong Rayleigh scattering seen at short wavelength,
though we slightly under-predict the observations at wavelengths
shorter then $\sim 0.6\mu m$. Our predicted emission spectrum agrees
remarkably well at $5.8$ and $8\mu m$, but slightly over-predicts the
emission at $3.6$ and $4.5\mu m$ when compared to the latest analysis
by \citet{knutson2012}. Our simulated IRAC phasecurves agree fairly
well with the amplitudes of variations, shape, and phases of minimum
and maximum flux. However, we over-predict the peak amplitude at
$3.6\mu m$ and $4.5\mu m$, and slightly under-predict the location of
the phasecurve maximum and minimum. These simulations include, for the
first time in a multi-dimensional simulation, a strong Rayleigh
scattering component to the absorption opacity, necessary to explain
observations in the optical and UV. The agreement between our models
and observations suggest that including the effects of condensates in
simulations as the dominant form of opacity will be very important in
future models.
\end{abstract}


\section{Introduction}
Among the $\sim 630$ known exoplanets, HD189733b currently boasts the
most numerous and comprehensive set of observations to date. A member
of the well know class of short-period, highly irradiated gaseous
planets (hot Jupiters) its relative proximity to earth and the
favorable star to planet radius ratio have made it the preferred
target for many different groups studying radial velocity, primary and
secondary transits, transit spectra, emission spectra, variability,
and phasecurves across a wide range of wavelengths. Given this wealth
of knowledge, it is the best candidate for detailed comparisons
between model predictions and observations. In this paper we compare
the results from our three-dimensional radiative-hydrodynamical
simulations of HD189733b with this wide array of observations.

Hot-Jupiters have been found around approximately $10\%$ of solar-type
stars. Their short orbital periods ($\sim 1-4$ days) suggests these
planets are subject to strong tidal forces that synchronize and
circularize their orbits. As a result one side perpetually faces its
host star, receiving intense irradiation, while the other side
receives no stellar irradiation. In the absence of any hydrodynamics,
the dayside reaches temperatures of several thousand degrees, while
the nightside temperature, maintained by the internal energy of the
planet, would remain at several hundred degrees. However, this large
temperature differential across the planet acts as an enormous driving
force for the atmospheric gas generating supersonic winds that advect
energy across the planet. The resulting temperature distribution
depends crucially on the non-linear coupling between the radiative
transfer of energy and hydrodynamic transfer of energy. The spatially
varying temperature across the planet leads to varying chemical
composition, and emission and absorption efficiencies and is thus
directly tied to observable properties. HD189733b is a fairly typical
example of such a planet.

There are a wide range of numerical approaches that have been taken to
address the coupled radiation hydrodynamics of irradiated gas
giants. Though coupled, it is convenient to separate a given studies
approach for the radiative transfer from that for the
hydrodynamics. Though not an exhaustive list, previous approaches to
radiative transfer include relaxation methods ({\it i.e.}  Newtonian
heating) \citep{showman2002, showman2008_2, cooper2005, cooper2006,
  langton2007, langton2008, menou2009, rauscher2010,perna2010},
kinematic constraints designed to represent incident flux
\citep{cho2003, cho2008, rauscher2008}, 3D flux-limited diffusion
(FLD) with \citep{dobbsdixon2010} and without
\citep{burkert2005,dobbsdixon2008} a separate radiation component,
dual grey 1D two-stream approximation
\citep{heng2011,rauscher2012,rauscher2012_2}, 3D FLD coupled to
wavelength dependent stellar irradiation \citep{dobbsdixon2012_1} and
1D frequency-dependent radial radiative transfer
\citep{showman2009}. Previous methods used to solve the hydrodynamical
portion include solving the equivalent barotropic equations
\citep{cho2003, cho2008, rauscher2008,rauscher2012,rauscher2012_2},
the shallow water equations \citep{langton2007, langton2008}, the
primitive equations \citep{showman2002, showman2008_2, showman2009,
  cooper2005, cooper2006, menou2009, perna2010, rauscher2010,
  heng2011}, Euler's equations \citep{burkert2005,dobbsdixon2008}, and
the Navier-Stokes equations \citep{dobbsdixon2010,
  dobbsdixon2012_1}. There are a wide range of assumptions in all
these approaches. Currently the most sophisticated approach to
hydrodynamics utilizes the Navier-Stokes equations, while that for
radiative transfer are 1D frequency-dependent models.

Given the wide range of approaches utilizing a wide range of numerical
codes, HD189733b provides an excellent target for not only testing how
well these models perform but also exploring the physical processes
included. Therefore, in this paper we present three-dimensional models
specifically tuned for HD189733b utilizing a model coupling wavelength
dependent two-stream approximation, wavelength dependent stellar
irradiation, and the fully compressible Navier-Stokes
equations. Section (2) presents the details on our modeling
methodology including both the dynamics and radiation in the
simulations and calculating observable signatures utilizing the models
results. Section (3) presents results illustrating both the
temperature and dynamical structure across the planet and detailed
comparisons to observations. Section (4) concludes with a discussion
of our results.

\section{Modeling Methodology}
In this paper we present results from a numerical model coupling the
fully compressible 3D Navier-Stokes equations to a two-stream,
frequency dependent radiative transfer calculation. Though we solve
the Navier-Stokes equations in a manner similar to previous papers
\citep{dobbsdixon2008,dobbsdixon2010,dobbsdixon2012_1} the radiative
transfer methodology is presented here for the first time. Observables
(phasecurves, spectra, etc.) are calculated utilizing the results of
the coupled radiative-hydrodynamical models. We describe all three
components (hydrodynamics, radiative transfer, and the calculation of
observables) in detail below.

\subsection{Hydrodynamics}
The hydrodynamical portion of the model solves the fully compressible
three-dimensional Navier-Stokes equations given by
\begin{eqnarray}
\pdt{{\bf u}} + \left({\bf u}\cdot\nabla\right) {\bf u} = -
\frac{\nabla{P}}{\rho} + {\bf g} -2{\bf \Omega\times u} \\ \nonumber - 
{\bf\Omega\times}\left({\bf\Omega\times r}\right) + \nu\nabla^2{\bf u}
+\frac{\nu}{3}\nabla\left(\nabla\cdot{\bf u}\right)
\label{eq:momentum}
\end{eqnarray}

\begin{equation}
\pdt{\rho} + \nabla\cdot\left(\rho{\bf u}\right) = 0.
\label{eq:continuity}
\end{equation}
\begin{equation}
\left[ \pdt{\epsilon} + ({\bf u}\cdot\nabla) \epsilon \right] = - P \,
\nabla \cdot {\bf u} - \nabla_{r}{\bf F_R} + S_{\star} + D_\nu,
\label{eq:thermalenergy}
\end{equation}
representing the momentum, continuity, and thermal energy equations
respectively. The three-dimensional velocity is given by {\bf u},
$\rho$ and $P$ are the gas density and pressure, and $\Omega$ is the
rotation frequency. The gravitational acceleration, ${\bf g}$ is taken
to be constant and purely radial. $\nu=\eta/\rho$ is the constant
kinematic viscosity. We have neglected the coefficient for the bulk
viscosity. The thermal energy equation contains heating and cooling
contributions from the radial gradient of the radiative flux ${\bf
  F_{R}}$, the incident stellar energy $S_{\star}$, and the viscous
heating $D_{\nu}$. The formalism for ${\bf F_{R}}$ and $S_{\star}$ is
detailed in Section (\ref{sec:radtransfer}) and that for $D_{\nu}$ can
be found in \citet{kley1987}.

We solve these equations in spherical coordinates with a resolution of
$\left(N_{r},N_{\phi},N_{\theta}\right)= \left(100,160,64\right)$
where $\phi$ is the longitude and $\theta$ the latitude. At the
equator this corresponds to cells that are $2.25^{\circ}$ by
$3^{\circ}$. Flow over the pole is accounted for via the method
described in \citet{dobbsdixon2012_1}.

\subsection{Radiative Transfer}
\label{sec:radtransfer}
In highly irradiated planets, transfer of energy via radiation plays a
very important role. For HD189733b this radiation can be broken into
two distinct contributions: the irradiation the planet receives
directly its host star and the local radiation comprised of
re-radiated stellar energy and the flux from the planets
interior. Though we include a full wavelength dependent treatment, the
incoming stellar irradiation can be largely characterized as short
wavelength (visible) and the re-radiated characterized as long
wavelength (infrared).

\subsubsection{Two-Stream Approximation}
To address radiative transfer of the re-radiated energy we have
developed a {\it frequency dependent}, two-stream approximation
\citep{mihalas1978} for the radial radiative flux, separating it into
multiple upward ($\fup$) and downward ($\fdown$) propagating
channels. The governing equations in each wavelength bin can be
written as
\begin{equation}
\frac{d\fup}{d\tau_{b}} = \fup - S_{b},
\label{eq:fup}
\end{equation}
\begin{equation}
\frac{d\fdown}{d\tau_{b}} = -\fdown + S_{b},
\label{eq:fdown}
\end{equation}

We have subdivided the full spectrum into $30$ wavelength bins (identical
to those in \citet{showman2009,dobbsdixon2012_1}). Assuming the gas
emits as a blackbody, the source function $S_{b}$ in Equations
(\ref{eq:fup}) and (\ref{eq:fdown}) in a given bin can be written as
\begin{equation}
S_b = \pi \int_{\nu_1\left[b\right]}^{\nu_2\left[b\right]}
B_{\nu}\left(T,\nu\right)d\nu 
\end{equation}

These equations require two boundary conditions. For these we choose
$\fdown\left(r=R_p\right)=0$ at the surface and
$F_{\uparrow,b}\left(r=R_{bot}\right)= S_b\left(T_{base}\right)$ at
the interior, where we fix the total upward flux
\begin{equation}
\displaystyle\sum\limits_{b} S_b\left(T_{bot}-\frac{dT}{dr}\Delta
r\right)\Delta r = F_{int}
\end{equation}
to the value required to match the observed radius. Note that though
we set the downward flux to be zero at $r=R_{p}$ in Equation
(\ref{eq:fdown}), stellar heating is accounted for directly in the
thermal energy equation (Equation (\ref{eq:thermalenergy})). This term
is discussed further in the next subsection. The optical depth in
Equations (\ref{eq:fup}) and (\ref{eq:fdown}) are calculated utilizing
the averaged opacity within that bin and a diffusivity factor,
\begin{equation}
\tau_b\left(r\right) = -1.66\int^{r}_{R_p}\rho\kappa_b dl.
\end{equation}
The diffusivity factor of $1.66$ is an approximation that accounts for
an exponential integral that arises when taking the first moment of
the intensity to calculate the flux \citep{elsasser1942}. The opacity
$\kappa_b$ is discussed in more detail below.

We solve Equations (\ref{eq:fup}) and (\ref{eq:fdown}) in the standard
way, utilizing an integrating factor $e^{\tau}$. Together with our
boundary conditions we find
\begin{equation}
\fdown \left(\tau_b\right) =
\int_0^{\tau_b}S_be^{-\left(\tau_b-\tau_b^{\prime}\right)}
d\tau_b^{\prime}.
\label{eq:fdownsolution}
\end{equation}
Note that if we wished to include the stellar heating directly into
these equations there would be an additional exponentially attenuated
term $S_{b,\star}e^{-\tau_b}$ added to the above equation for
$\fdown$, with
\begin{equation}
S_{b,\star} = \left(\frac{R_{\star}}{a}\right)^2\pi
\int_{\nu_1\left[b\right]}^{\nu_2\left[b\right]}
B_{\nu}\left(T_{\star},\nu\right)d\nu
\end{equation}

The equation for the upward flux does include an initial source term
due to the flux from the interior, and can be written as
\begin{equation}
\fup\left(\tau_b\right) =
S_{b}\left(T_{bot}\right)e^{-\left(\tau_{b,bot}-\tau_b\right)} +
\int_{\tau_{b,bot}}^{\tau_b}S_be^{-\left(\tau_b^{\prime}-\tau_{b}\right)}
d\tau_b^{\prime}.
\label{eq:fupsolution}
\end{equation}
Finally, the net radial flux (defined with positive indicating outward
flux) can be computed as
\begin{equation}
F_{r}\left(\tau_b\right) = \displaystyle\sum\limits_{b}
\left(\fup\left(\tau_b\right)-\fdown\left(\tau_b\right)\right).
\label{eq:ftotal}
\end{equation}

However, it is important to note that Equation (\ref{eq:ftotal}) only
works in the upper atmosphere. As demonstrated by
\citet{rauscher2012}, in regions below the photosphere the
temperature-pressure profile becomes isothermal. As the density
increases, the integrated optical depth across a grid cell becomes
very large, and the profile becomes dependent on the numerical
resolution. The resolution required to compute an accurate temperature
profile at depth using Equation (\ref{eq:ftotal}) far exceeds
computational capabilities. Results from other groups that utilize
some form of two-stream approximation \citep{showman2009,heng2011}
seem to contain similar isothermal regions at depth, though they do
not discuss this issue. To surmount this issue, we follow
\citet{rauscher2012} and transition to a diffusive scheme at
$\tau=2.5$. This is done in each wavelength bin independently given
that the photosphere can be at significantly different depths across
the spectrum. Given the larger optical depths, a diffusion
approximation is entirely adequate in the deeper atmosphere. Below the
active weather layer, radial radiative diffusion remains the dominant
form of energy transport all the way down to the radiative-convective
boundary. Finally, as seen in \citet{dobbsdixon2010}, non-radial
radiative transfer does play a role in influencing the
energetics. Though inconsequential near the sub-stellar point, the
horizontal flux of radiation can play an important role. We have
neglected the non-radial contribution here.

\subsubsection{Stellar Heating}
\label{sec:stellarheating}
The irradiation of the planets atmosphere from the central star is
directly accounted for in Equation (\ref{eq:thermalenergy}) for the
thermal energy of the gas. We choose this procedure rather then
inserting an additional boundary term into Equation
(\ref{eq:fdownsolution}) for $\fdown$ as it more accurately captures
the slant-optical depth associated with the path of stellar photons
through the atmosphere. The formalism here is identical to the
procedure used in \citet{dobbsdixon2012_1}. The stellar heating term
in the thermal energy equation can be written as
\begin{equation}
S_{\star}= \left(\frac{R_{\star}}{a}\right)^2
\displaystyle\sum\limits_{b=1}^{nb}\frac{d\tau_{b,\star}}{dr}
e^{-\tau_{b,\star}/\mu_{\star}}\int_{\nu_{b,1}}^{\nu_{b,2}}\pi
B_{\nu}\left(T_{\star},\nu\right)d\nu,
\label{eq:stellarheat}
\end{equation}
where $B_{\nu}\left(T_{\star},\nu\right)$ is the stellar blackbody for
HD189733A taken from the Kurucz models\footnote[1]{see
  http://kurucz.harvard.edu/stars.html}. $R_{\star}$ and $a$ represent
the stellar radius and semi-major axis. The slant-optical depth is
accounted for by including $\mu_{\star}$, the cosine of the angle
between the normal and the incident stellar photons. The optical depth
to stellar photons is calculated as
\begin{equation}
  \tau_{b,\star} = \int\rho\kappa_{b,\star}dl.
\label{eq:taustar}
\end{equation}

\subsubsection{Opacities}
The details of the atmospheric opacity is a crucial parameter in
determining the energy deposition and re-radiation, subsequently
playing an important role in the overall atmospheric
dynamics. \citet{dobbsdixon2008} explored the role of changing the
opacity within the context of a grey flux-limited diffusion (FLD)
approximation. The results of this study indicated that as opacity of
the atmosphere is decreased, stellar energy is deposited at larger
pressures, where the dynamics is more effective at redistributing it
to the nightside. In this paper we are utilizing a much more
sophisticated radiative transfer routine, but the general principle
still holds, increasing opacities causes the energy to be deposited
much higher in the atmosphere.

Here we are utilizing the frequency dependent opacities $\kappa_{\nu}$
of \citet{sharp2007} which assume solar composition gas. The dominant
molecules relevant for HD189733b in these calculations are water,
methane and carbon monoxide. The frequency dependent opacities are
averaged within each wavelength bin utilizing a Planck mean,
\begin{equation}
\kappa_b = \frac{\int_{\nu_1\left[b\right]}^{\nu_2\left[b\right]}
  \kappa_{\nu}\left(\rho,T\right)B_{\nu}\left(T\right)d\nu}
      {\int_{\nu_1\left[b\right]}^{\nu_2\left[b\right]}B_{\nu}\left(T\right)d\nu}.
\label{eq:kappab}
\end{equation}
However, we find that the exact form of the averaging method does not
have a significant effect on the radiative solution. For the
re-radiated component of radiation (Equations (\ref{eq:fdownsolution})
and (\ref{eq:fupsolution})) the local gas temperature is used in
$B_{\nu}\left(T\right)$. When computing the contribution from the
incident stellar irradiation (Equation (\ref{eq:taustar})) we
replace this with the stellar blackbody,
$B_{\nu}\left(T_{\star}\right)$.

After significant experimentation, we found that
radiative-hydrodynamical models utilizing simply the opacities
directly from \citet{sharp2007} did not agree well with the
observations. Comparison to emission spectra
\citep{deming2006,knutson2007_2,knutson2009,charbonneau2008,
  knutson2012} indicated these initial models were too cool in the
upper atmosphere, and comparison to transit spectra
\citep{pont2008,sing2011,pont2012} indicated that the existing
Rayleigh contribution in the opacities was woefully inadequate to
explain the strongly varying transit radius with decreasing
wavelength. To address this issue we have supplemented the opacities
with two addition components, a uniform grey component $\kappa_{grey}$
at all wavelengths and an component that scales as $\lambda^{-4}$, as
would be expected from Rayleigh scattering. This extra opacity can be
written as
\begin{equation}
  \kape\left(\lambda\right) = \kappa_{grey} +
  \kappa_{RS}\left(\frac{\lambda}{0.9\mu m}\right)^{-4}.
\end{equation}
We have run a number of simulations varying the magnitude of these two
components. Setting $\kappa_{grey}=0.035 cm^2/g$ and $\kappa_{RS}=0.6
cm^2/g$ provides the best fits to the observations. Note that the
magnitude of the grey component is the same as that required in the
one-dimensional models in \citet{knutson2012}. This extra opacity is
added self-consistently to all the opacities utilized in the code and
in the post-processing described in the next sub-section. This is the
first multi-dimensional radiative hydrodynamical model to include such
a strong Rayleigh scattering component. Note however that we are
treating $\kape$ solely as an absorptive opacity, ignoring its
scattering properties. This should be included in future work.

\subsection{Calculating Observables from 3D Models}
\label{sec:calcobs}
To determine the transmission spectrum from the hydrodynamical models,
we calculate the wavelength dependent absorption of stellar light
traversing through the limb of the planet. This allows us to determine
the effective radius of the planet and a fractional reduction of
stellar flux $F_{\star}$. Neglecting limb-darkening of the star, this
can be expressed as
\begin{equation}
\left(\frac{F_{in transit}}{F_{\star}}\right)_{\lambda} = \frac{\int
  \left(1-e^{\tau\left(b,\alpha,\lambda\right)}\right)b db d\alpha}{\pi
  R_{\star}^{2}},
\end{equation}
where $\tau\left(b,\alpha,\lambda\right)$ is the total optical depth
along a given chord with impact parameter b and polar angle $\alpha$,
defined on the observed planetary disk during transit.

To calculate the emission spectrum and phasecurves, we integrate
inwards along rays parallel to the line of sight. The emerging
Blackbody flux from the planet at each point on the surface is given
by
\begin{equation}
 B\left(x,y,\lambda\right) = \int_{0}^{\infty}\frac{2\pi
    hc^2/\lambda^5}{exp\left(\frac{hc}{\lambda kT}\right)-1}e^{-\tau}d\tau.
 \label{eq:emerging}
\end{equation}
To calculate the apparent dayside luminosity, we integrate over the
observed disk
\begin{equation}
  L_p\left(\lambda\right) = \int I_{p}dA,
\end{equation}
where $dA$ is taken over the observer's plane and the emerging
intensity is given by
\begin{equation}
I_{p}\left(x,y,\lambda\right) = \frac{\lambda^2}{c} \int
B_{\lambda}\left(x,y,\tau\right)e^{-\tau_{\lambda}}d\tau_{\lambda}.
\label{eq:intensity}
\end{equation}
For both transmission
and emission calculations, the density and temperature needed to
calculate $\tau$ at each location are interpolated from the values in
the 3D models.

\section{Results}
In this section we present the results of our 3D radiative
hydrodynamical simulations of the irradiated planet HD189733b. We
first discuss the wind and temperature structures throughout the
atmosphere and then compare synthetic observations calculated using
these results to a wide variety of observations. 

\subsection{Temperature and Wind Structure}
As discussed in Section (1) the synchronous rotation and intense
irradiation of the dayside by the host star HD189733A drives winds
throughout the upper portions of HD189733b. These winds advect energy
throughout the atmosphere resulting in significantly different
temperature distributions then expected from purely radiative
calculations. As is to be expected, the details of the dynamics
depends on the input parameters; Table (\ref{table:one}) lists a
number of these parameters.

\begin{table}
\begin{center}
\begin{tabular}{|c|c|}
\hline
Parameter & Value \\
\hline
$P_{rot}$, $P_{orb}$ & $2.22$ days \\
\hline
$R_{\star}$ & $0.76 R_{\odot}$ \\
\hline
$R_{p}$ & $1.138 R_{Jup}$ \\
\hline
$\Mp$ & $1.144 M_{Jup}$ \\
\hline
a & $0.031$AU \\
\hline
$\nu$ & $10^7 cm^2 s^{-1}$ \\
\hline
\end{tabular}
\end{center}
\caption{Physical parameters chosen for simulations of HD189733b
  presented in this paper taken from \citet{torres2008}.}
\label{table:one}
\end{table}

\begin{figure}
  \centering
  \includegraphics[width=1\linewidth]{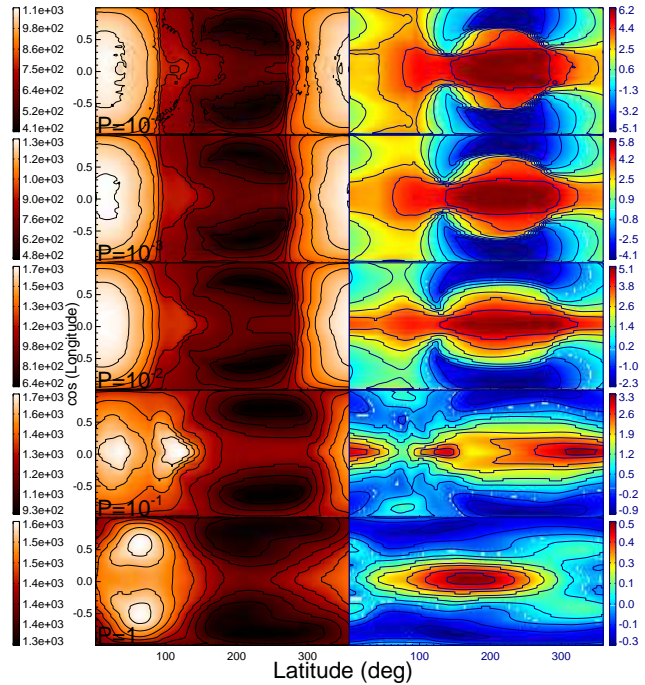}
  \caption{Temperature in Kelvin (left) and zonal speed in km/s
    (right) at $10^{-4}$, $10^{-3}$, $10^{-2}$, $10^{-1}$, and $1$
    bars from top bottom. The images are projected onto a {\it
      cos}(longitude) map to highlight the importance of the
    equatorial features in determining observable properties. The
    sub-stellar point is located at $0^{\circ}$ latitude and the
    equator runs horizontally through the center of each plot.}
  \label{fig:tvel}
\end{figure}

The radial distribution of the energy deposited in the atmosphere from
its host star extends from the upper regions of the atmosphere
($10^{-5}$ bars) down to $\sim 10$ bars due to the complicated
wavelength dependent opacities. Coupled with radial mixing, this leads
to varying temperature distributions at different pressure levels
within the atmosphere. Figure (\ref{fig:tvel}) illustrates the
temperature and zonal (i.e. longitudinal) velocity at constant
pressure surfaces ranging from $10^{-4}$ to $1$ bars. There are a
number of features of interest in these plots. The hottest point in
the atmosphere is near the sub-stellar point just above $\sim 0.1$
bars, corresponding to the location where the majority of the stellar
energy is deposited. The coolest points are high in the atmosphere in
the nightside gyres found at mid-latitudes. Zonal jets, super-rotating
at the equator and counter-rotating at mid latitudes are also quite
evident. Jet velocities increase with decreasing atmospheric pressure,
reaching $6$ km/s at pressures of $10^{-4}$ bars. The advective
signature of the jets is seen near the terminators ($90^{\circ}$ and
$270^{\circ}$) as tongues of hotter regions stretching onto the
nightside. Deeper in the atmosphere, where the radiative time-scale is
longer, advection is more efficient, equalizing temperature and moving
the hottest point significantly downwind to the east of the sub-stellar
point.

\begin{figure}
  \centering
  \includegraphics[width=1.1\linewidth,trim=20mm 5mm 5mm 0mm,clip]{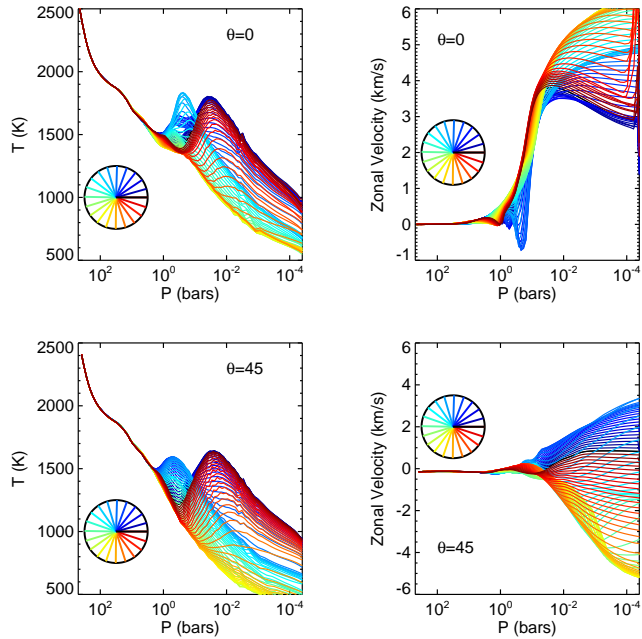}
  \caption{Temperature versus pressure (left panels) and zonal
    velocity versus pressure (right panels) at the equator and
    mid-latitudes. The inset color wheel indicates the longitude of a
    given profile. The star is located t the right of the inset color
    wheel, thus the black line denotes the sub-stellar longitude. The
    eastward (positive) direction is counter clockwise.}
\label{fig:pt}
\end{figure}

Figure (\ref{fig:pt}) illustrates the behavior of the temperature and
the zonal velocity with pressure at both the equator and
mid-latitudes. Several separate temperature peaks are seen in these
plots. The temperature inversion seen just below $\sim 10^{-2}$ bars
near the sub-solar point again corresponds with the maximum energy
deposition associated with the stellar heating. The larger
temperatures at longitudes between $90$ and $100^{\circ}$ and
pressures between $1$ and $10^{-1}$ bars corresponds to the heating
caused when the westward mid-latitude jet deviates toward the equator
and interacts with the eastward equatorial jet. At mid-latitude the
heating between $1$ and $10^{-1}$ bars is generated by the compression
and shock-heating associated with converging flows. Much of this
behavior can also be seen in the zonal velocity shown in Figure
(\ref{fig:tvel}). Deeper in the planet temperatures converge onto a
single curve that will continue to steepen until becoming unstable at
the radiative-convective boundary.

The zonal velocity is almost entirely eastward (super-rotating) at the
equator. The largest velocities are found high in the atmosphere on
the nightside of the planet just past the eastern terminator. These
flows have very high Mach numbers, reaching up to $3.5$ at the western
terminator where the flow has cooled radiatively but still retains
high velocities. With such high velocities it is likely that shocks
play an important role in limiting the velocity. Though our numerical
scheme can capture larger scale shocks, it is likely there are
additional processes occurring on size scales smaller then our grid
that contribute to the overall drag of the flow and that these
velocities represent upper limits to the actual flow speeds. At
mid-latitude, flow is in both the eastward and westward directions at
low pressure. As the pressure increases, circumplanetary westward jets
develop.

\subsection{Comparison to Observations}
HD198733b is currently the most observed extrasolar planet given its
relative proximity and favorable planet-star ratio. Therefore in this
section we perform detailed comparisons between these observations and
synthetic observations calculated utilizing our numerical model. The
methods for calculating observable metrics from the simulations were
presented in Section (\ref{sec:calcobs}).

\begin{figure}
  \centering
  \includegraphics[width=1.1\linewidth,trim=20mm 5mm 5mm 0mm,clip]{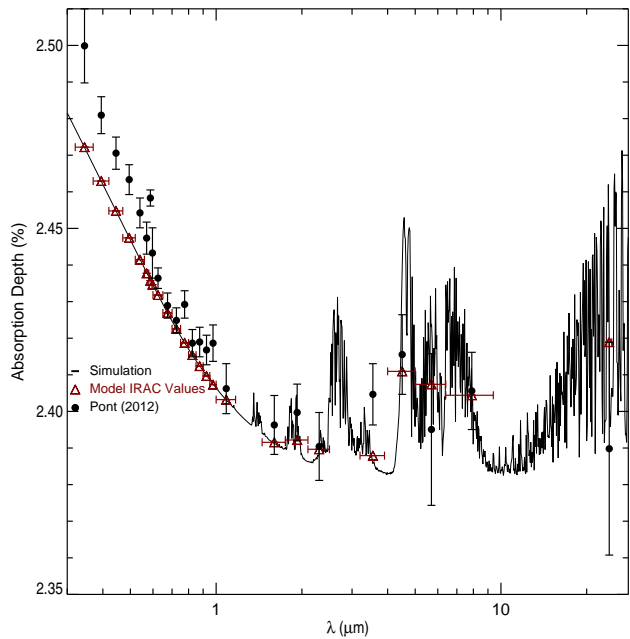}
  \caption{The transit spectrum for the simulated planet
    HD198733b. Over-plotted are data from \citet{pont2012} as solid
    points, and the band averaged mode values as red triangles. Bands
    are taken from Table 6 of \citet{pont2012}. Both the data and model
    exhibit strong Rayleigh scattering at short wavelengths and muted
    molecular features at longer wavelengths.}
  \label{fig:transit}
\end{figure}

In Figure (\ref{fig:transit}) we present the calculated transit
spectrum from the UV through the infrared. The calculated transit
spectrum is dominated by Rayleigh scattering below $\sim 1\mu m$ with
molecular features becoming important at longer wavelengths. However,
the uniform component of $\kape$ that we have added to the opacity
suppresses the strength of many of these features. The observational
data shown in Figure (\ref{fig:transit}) is a reanalysis of data from
a number of groups by \citet{pont2012}. Absorption depths reported by
a number of groups exhibit significant scatter \citep{pont2008,
  beaulieu2008, sing2009, sing2011, gibson2012, desert2011,
  desert2009, knutson2007_2, knutson2009,
  ehrenreich2007}. \citet{desert2011} and \citet{pont2012} suggest
this may be due to stellar variability. \citet{pont2012} has recently
reanalyzed much of this data utilizing a consistent treatment for
stellar spot corrections, transit properties, and stellar-limb
darkening. Our model agrees fairly well with these re-analyzed
data. We do see deviations at very short wavelengths ($<0.6\mu m$) in
the Rayleigh scattering portion of the spectrum. This is likely due to
either a slight temperature inversion in the uppermost portion of the
atmosphere that our model is not capturing or a change in grain size
with height, physics not included in our current model. The very
uppermost radius of our model corresponds to an absorption depth of
$\sim 2.45\%$ and pressures of several microbars. We attempt to account
for any additional absorption by adding an isothermal region above
this pressure with the temperature within each radial column taken
from the simulated region. The limits on the radial extent is a
numerical constraint associated with very low density in this region
and it may play a role in the deviations we see at wavelengths
$<0.6\mu m$.

\begin{figure}
  \centering
  \includegraphics[width=1.1\linewidth,trim=20mm 5mm 5mm 0mm,clip]{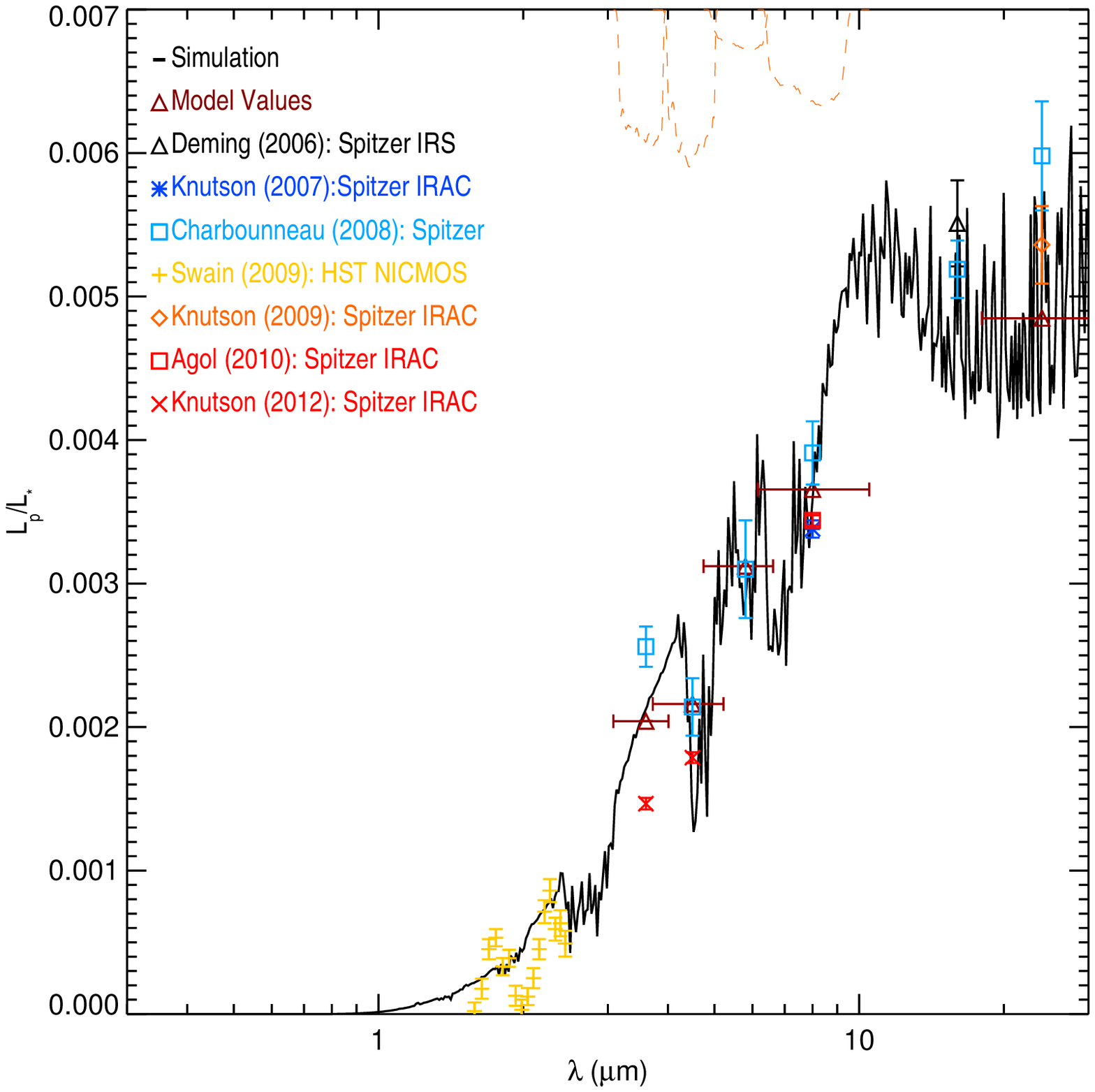}
  \caption{The dayside emission spectrum for HD198733b shown as the black
    curve. Over-plotted are data from \citet{deming2006, swain2009,
      knutson2007_2, charbonneau2008, knutson2009, knutson2012}}
  \label{fig:emission}
\end{figure}

\begin{figure}
  \centering
  \includegraphics[width=1.1\linewidth,trim=20mm 5mm 5mm 0mm,clip]{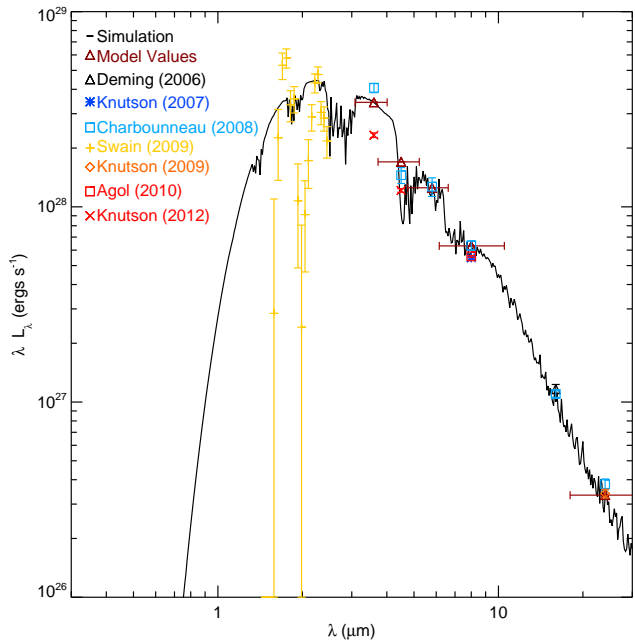}
  \caption{Apparent dayside luminosity $\lambda L_{\lambda}$ during
    secondary transit clearly illustrating the wavelengths where a vast
    majority of the radiative energy that ultimately drives the dynamics
    is found.}
  \label{fig:lambda_L}
\end{figure}

\begin{figure}
  \centering
  \includegraphics[width=1.03\linewidth]{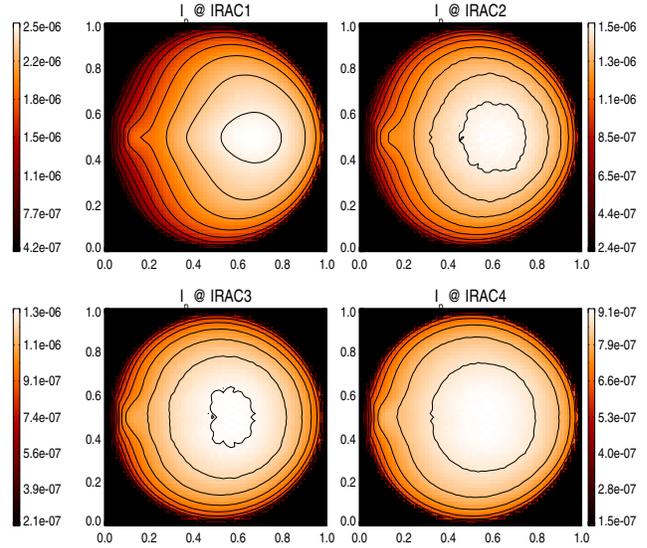}
  \caption{The resolved emerging intensity given by Equation
    (\ref{eq:intensity}) with units of $ergs
    cm^{-1}str^{-1}cm^{-2}s^{-1}$, during secondary eclipse as would
    it would be seen in the four IRAC bands across the entire face of
    the planet. The sub-stellar point lies in the center of each
    image. The importance of the circumplanetary jet and the varying
    degrees of asymmetry are evident.}
  \label{fig:iracmap}
\end{figure}

The next diagnostic we examine is the emerging spectrum on the dayside
of the planet. Figure (\ref{fig:emission}) illustrates the calculated
emission spectrum during the secondary eclipse. Again, we find general
agreement throughout the infrared, slightly over-predicting the newest
measurements at $3.6$ and $4.5\mu m$ from \citet{knutson2012}, but
doing very well at $5.8$ and $8\mu m$. We under-predict the flux from
\citet{knutson2009} MIPS measurement at $24\mu m$. We do not reproduce
the detailed shape seen by \citet{swain2009} at $\sim 2\mu m$. As our
overall goal is to utilize observations to validate our
radiative-hydrodynamical model we plot the actual emerging intensity
as $\lambda L_{\lambda}$ in Figure (\ref{fig:lambda_L}). From this
figure it is clear that the vast majority of the radiative energetics
is occurring in the region from $\approx 1-10\mu m$. Outside this
window the emerging intensity is orders of magnitude lower. The
dynamics (jet formation, advection, meridional circulation, etc.) will
overwhelmingly be driven by radiative energy transfer in this
window. Therefore, the agreement between our model and observations in
the IRAC bandpasses gives us confidence that we are capturing a
majority of the radiative energetics, despite some disagreement at
longer wavelengths. To get a better sense of the planetary emission in
the IRAC bandpasses, we plot the spatially resolved emission from the
planet in Figure (\ref{fig:iracmap}) during the center of the
secondary eclipse. These figures illustrate a number of features
including the offset of the hot-spot, the advection of energy by the
super-rotating equatorial jet, and the varying level of asymmetry in
different bandpasses, indicative of the varying pressure levels they
probe. The IRAC 4 image compares favorably to the 2D secondary
eclipse map of \citet{majeau2012,dewit2012} utilizing $8\mu m$ data.

\begin{figure}
  \centering
  \includegraphics[width=1.1\linewidth,trim=20mm 1mm 0mm 0mm,clip]{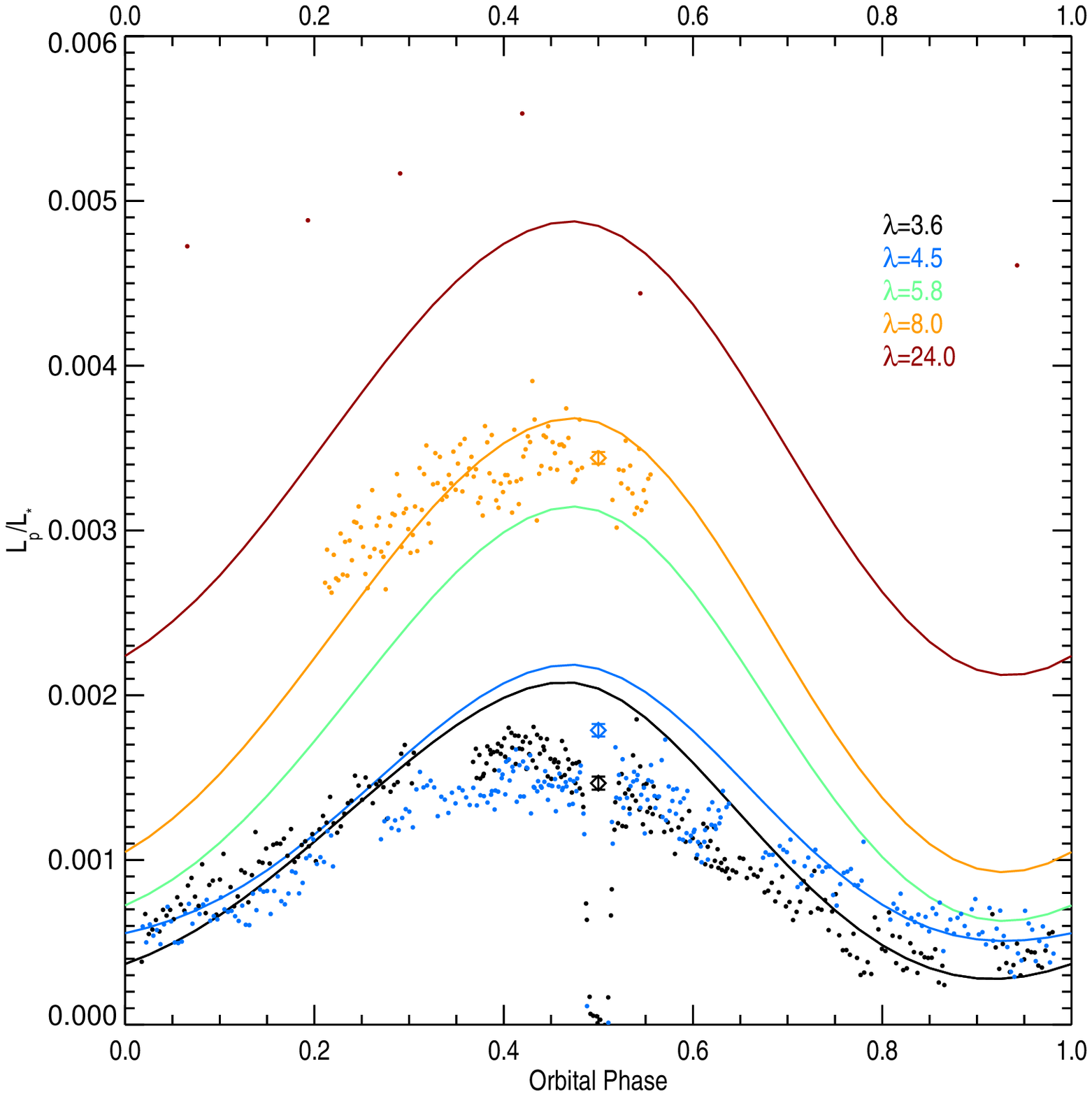}
  \caption{The simulated IRAC phasecurves for HD189733b.  Over-plotted
    are observed phasecurves from \citet{knutson2009} at $24\mu m$,
    \citet{agol2009} at $8\mu m$, and \citet{knutson2012} at $4.5$ and
    $3.6\mu m$.}
  \label{fig:phase}
\end{figure}

One of the most important diagnostics of atmospheric dynamics on
extrasolar planets are multi-wavelength phasecurves of the planet
throughout its entire orbit. As varying phases of the planet come into
view, we are able to directly compare the observed flux to that
predicted by models. Deviations from radiative equilibrium indicates
advective contributions to energy redistribution via the atmospheric
dynamics. The presence of a strong circumplanetary jet was first
robustly demonstrated for the planet HD209458b
\citep{knutson2007_2}. Therefore, we examine the multi-wavelength
phasecurves. The amplitude, shape, and the phases of the the minimum
and maximum of the flux are all important diagnostics. In
Figure(\ref{fig:phase}) we show the simulated phasecurves in the IRAC
and MIPS bandpasses compared to observations from
\citet{knutson2009,agol2009,knutson2012}. In general we find
remarkable agreement between our models and the data. We do very well
in matching the observations of \citet{agol2009} at $8\mu m$,
including its distinct, non-symmetric shape. At $3.6\mu m$ we slightly
over-predict the flux at secondary eclipse (orbital phase of $0.5$),
and very slightly under-predict the flux from the nightside. At
$4.5\mu m$ we also over-predict the flux at secondary eclipse, but
match the nightside flux fairly well. Finally we significantly
under-predict the flux at $24\mu m$. We saw this behavior in Figure
(\ref{fig:emission}) as well, but as discussed above the results at
$24\mu m$ likely have little influence on the overall atmospheric
dynamics. However, in comparing our model to the phasecurve
observations it is important to note that \citet{pont2012} claims that
\citet{knutson2012} overstates the ability of Gaussian Processing to
correct for variability on the 1-2 day time-scale. Thus the errors in
their phasecurve amplitude are likely $\sim 5$ times too small. This
suggests our amplitudes may consistent with the data given these
larger uncertainties. Also seen in Figure(\ref{fig:phase}) is the
distinct offset of the phase of maximum and minimum fluxes from $0.5$
and $0$, respectively. Our models agree very well with the shape of
the curve in the IRAC wavelengths of $3.6\mu m$, $4.5\mu m$ and $8\mu
m$. The observations at $24\mu m$ are too sparsely sampled to
definitively compare the phase-offsets. Table (\ref{table:two}) lists
the observed and simulated phase difference from transit and secondary
eclipse for the minimum and maximum fluxes respectively. Comparison
between the observations and the models shows that we systematically
under-predict the magnitude of the offset at all wavelengths, save an
over-prediction of the minimum at $4.5\mu m$. Taken at face-value this
suggests that the jet strength may be somewhat underestimated in our
models.

\begin{table*}
  \begin{minipage}{126mm}
  \begin{center}
    \caption{Comparison between the calculated and observed offsets of
      minimum and maximum fluxes (in hours) from the inferior and superior
      conjunction, respectively, in the IRAC and MIPS bandpasses. Observed
      values are taken from \citet{knutson2012} Table 2. Note that
      duration and accuracy of the $8 \mu m$ and $24\mu m$ data was not
      sufficient define the minimum of the phasecurve in these wavebands.}
      \begin{tabular}{|c|c|c|c|c|}
        \hline
        $\lambda$ & Observed Max Offset & Calculated Max Offset & Observed Min Offset & Calculated Min Offset \\
        \hline
        $3.6 \mu m$ & $5.29\pm 0.59$ & 2.30 & $6.43\pm 0.82$ & $4.17$ \\
        \hline
        $4.5 \mu m$ & $2.98\pm 0.82$ & 1.82 & $1.37\pm 1.00$ & $3.20$ \\
        \hline
        $8.0 \mu m$ & $3.5\pm 0.4$ & 1.35 & --- & $3.49$ \\
        \hline
        $24.0 \mu m$ & $5.5\pm 1.2$ & 1.33 & --- & $3.20$ \\
        \hline
      \end{tabular}
  \end{center}
  \end{minipage}
  \label{table:two}
\end{table*}

\begin{figure}
  \centering
  \includegraphics[width=1.0\linewidth,trim=18mm 0mm 5mm 0mm,clip]{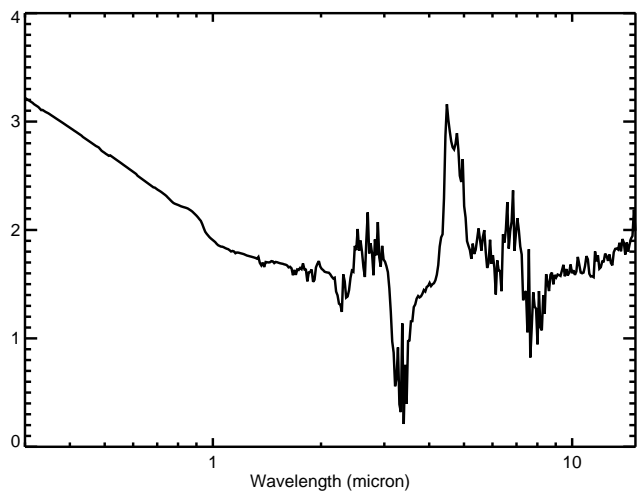}
  \caption{The effective time-offset as a function of
    wavelength. Timing variations are primarily due to temperature
    differences between the eastern and western hemispheres.}
  \label{fig:timeoffset}
\end{figure}

Though phasecurves provide an excellent diagnostic of the jet
dynamics, \citet{dobbsdixon2012_1} suggested a new observational
diagnostic to determine the strength of the jet. In the presence of a
strong jet the temperature at the eastern terminator (in the direction
of the jet) will be somewhat larger then that at the western
terminator. The differential scale-heights between the two terminators
will alter the timing of ingress and egress at a measurable
level. This effect will change strength at different wavelengths as
the day/night temperature differential depends on pressure and may be
detectable with simultaneous observations at multiple wavelengths. In
Figure (\ref{fig:timeoffset}) we present the apparent offset in the
time of central transit when the lightcurves computed from our model
are fit with a spherical planetary transit model (from
\citet{mandel2002}). We predict that the magnitude of this effect will
be most visible in observations comparing the $3.6$ and $4.5 \mu m$
regions of the spectra (with timing differences reaching up to 3
seconds), and also possibly at short wavelengths.

The final diagnostic we explore is the stability of the atmospheric
dynamics and thus the temperature distribution throughout the
atmosphere. \citet{agol2010} monitored secondary eclipse depths of $7$
events spread over $270$ days. They find very little change in the
observed depths, putting an upper limit on the variability of the
dayside flux of $2.7\%$. To test this, we have calculated the
secondary eclipse depth variation over $30$ orbits. Note that we don't
actually have to simulate $30$ orbits to derive this quantity. Because
we don't have observational constraints for our models we have simply
continuously monitored the expected depth over this period. We find
that the dayside emission from our models is extremely stable, with
RMS deviations of $0.1\%$, $0.07\%$, $0.07\%$, $0.07\%$, and $0.06\%$
at $3.6\mu m$ $4.5\mu m$, $5.8\mu m$, $8\mu m$, and $24\mu m$
respectively.

\begin{figure}
  \centering
  \includegraphics[width=1.1\linewidth,trim=20mm 1mm 0mm 0mm,clip]{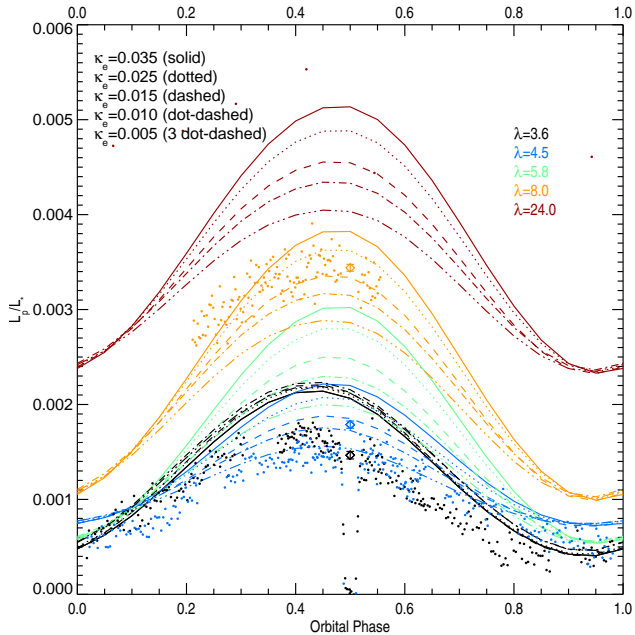}
  \caption{The simulated phase curves for the four IRAC bands and the
    MIPS band for simulations with a varying strength of the extra grey
    absorber $\kape$.}
  \label{fig:multiphase}
\end{figure}

In summary, our model shows fairly good agreement with all the
observations. It is possible that the jet is somewhat stronger then we
predict and thus more efficient at cooling the dayside and heating the
nightside, as indicated primarily by the $3.6\mu m$ phasecurve. In
addition, some fraction of the energy we see emitted at shorter
wavelengths is actually escaping at longer wavelengths. Given the
results presented in Figure (\ref{fig:lambda_L}) only a very small
fraction of the energy at short wavelengths must be transferred to
longer wavelengths to explain the observations. There are a number of
possible reasons that we are finding a discrepancy between the model
and the observations. One likely culprit is the composition and thus
the opacity of the atmospheric gas. As mentioned we are utilizing
solar composition, molecular opacities from \citet{sharp2007}
augmented with a extra wavelength dependent opacity of $\kape=0.035 +
0.6\left( \frac{\lambda}{0.9\mu m} \right)^{-4}$. In
Figure(\ref{fig:multiphase}) we illustrate the effect of varying the
strength of a simple grey component ($\kape$) on the
phasecurves. Indeed it is clear that increasing $\kape$ causes the
overall emission to shift from the $3.6\mu m$ band to longer
wavelengths. A small decrease at $3.6 \mu m$ is more then sufficient
to cause large shifts at other wavelengths.

\section{Discussion}
We have presented a new set of radiative hydrodynamical simulations
for the irradiated gas giant planet HD189733b. Simulations utilize a
frequency dependent two-stream approximation to calculate radiative
transfer coupled to a 3D, fully compressible solution of the
Navier-Stokes equations. Opacities, crucial for understanding the
deposition of stellar energy and the subsequent re-radiation and
cooling of the atmosphere, are calculated assuming contributions from
solar-composition molecules, a wavelength independent grey component,
and strong Rayleigh scattering. The latter two are assumed to come
from the presence of grains in the atmosphere.

The atmospheric dynamics is characterized by supersonic winds that
fairly efficiently advect energy from the dayside to the
nightside. Super-rotating equatorial jets form for a wide range of
pressures from $10^{-5}$ bars down to $\sim 10$ bars. Counter rotating
jets form at higher latitudes flanking the equator setting up a
pattern of three jets from north to south pole across the
nightside. The westward, counter rotating jets become slower at depth,
but are also able to circumnavigate the planet below pressure of $\sim
10^{-2}$ bars. Not only is the temperature structure modified by
advection of energy in these jets, but the interaction between the
jets also leads to compressional and shock heating, further modifying
the temperature distribution across the planet.

We performed detailed comparisons between our model and a number of
observations including transit spectra, emission spectra, eclipse
maps, phasecurves, and variability. Model transit spectra agree fairly
well with the data from the infrared to the UV, though they
under-predict the observations at wavelengths less then $\sim 0.6\mu
m$. This may be due to an increase in temperature or a change in grain
size, neither of which is captured by our models. Our predicted
emission spectrum agrees remarkably well at $5.8 \mu m$ and $8\mu m$, but
slightly over-predicts the emission at $3.6\mu m$ and $4.5\mu m$ when
compared to the latest analysis by \citet{knutson2012}. Our
measurement of the phasecurve at $8\mu m$ agrees very well,
reproducing both the peak amplitude and the phase of maximum flux. As
with the emission spectrum, we over predict the peak amplitude at
$3.6\mu m$ and $4.5\mu m$, though agree fairly well with the phase
offsets of minimum and maximum fluxes.

Among the other groups studying multi-dimensional radiation
hydrodynamics of irradiated hot Jupiter atmospheres,
\citet{showman2009}, \citet{fortney2010}, and \citet{knutson2012} have
also performed detailed comparisons to HD189733b data utilizing the
models of \citet{showman2009}. As both their model and opacities
differ from ours it is not unexpected that we differ in some of the
details. However, despite these differences we do appear to present a
broadly consistent picture of the dynamics. Perhaps the most important
difference between our results, with respect to the dynamics, is the
location of the hotspot, indicative of the strength of the winds and
the efficiency of advection. \citet{showman2009} significantly
over-predict this efficiency, finding the location of the hottest
point somewhat further downwind then observed. As a result of
increased advection the nightside temperatures they predict are larger
then observations. To address the discrepancy between the observed
phase of the hotspot and that of the model, \citet{showman2009} ran
additional models with metallicity enhanced $5$ times relative to solar
(at odds with the measured sub-stellar metallicity \citep{torres2008})
or a rotation period twice that of the orbital period (at odds with
tidal synchronization theory \citep[e.g.]{rasio1996_2}). Our fits do
not rely on these mechanisms, but rather an additional opacity source
most likely associated with condensates.

In an attempt to understand the discrepancy between their model
predictions and the data at $4.5 \mu m$ \citet{knutson2012} suggest
that advection of $CO$ from the dayside to the nightside (where the
Carbon should convert to $CH_4$ in chemical equilibrium) is
important. Simulations of \citet{cooper2006} and \citet{agundez2012},
studying the carbon distribution, support this claim. Give the
agreement between our simulations and the observations at $4.5\mu m$
on the nightside, it appears that we do not require increased CO on
the nightside. Moreover, the decreased advective efficiencies we find
in our simulations would self-consistently predict less $CO$ on the
nightside as well.  At the important shorter wavelengths of $3.6\mu m$
and $4.5\mu m$, we over-predict dayside fluxes in comparison to
observations of \citet{knutson2012}. If the amount of $CH_4$ on the
dayside was increased above chemical equilibrium values, as predicted
by photochemical models \citep{visscher2011}, the flux at $3.6 \mu m$
would be suppressed bringing our models into better agreement. As
noted in \citet{knutson2012} this would also decrease the flux at $8$
and $24\mu m$. However, it is important to remember that the addition
of condensates may mask many details associated with non-equilibrium
molecular distributions.

From our analysis of HD189733b, it is clear that the hazes and/or
grains play a significant, perhaps dominant, role in determining the
overall opacity. \citet{pont2012} suggest molecular opacities may not
be relevant for HD189733b, and almost all the features of their data
can be explained simply by a combination of a cloud deck and Rayleigh
scattering dust grains. Thus it is prudent to consider if the
atmosphere can form grains and hazes. Ideally one would
self-consistently calculate the formation, destruction, advection, and
settling of grains and hazes. This is not currently included in our
model, but is an important direction for future research. The vertical
velocities in the upper radiative zones are relatively modest compared
to the zonal flows, but when coupled to the large scale-height of the
atmosphere, could lead to significant vertical mixing required to keep
grains at low pressures.

\section*{Acknowledgments}
 We thank David Catling and Ty Robinson for discussions regarding the
 diffusivity factor, Heather Knutson for providing observational
 phasecurves, and Adam Burrows for providing opacities. IDD was
 partially supported by the Carl Sagan Postdoctoral program and the
 NASA Astrobiology Institute's Virtual Planetary Laboratory Lead Team,
 supported by NASA under solicitation NNH05ZDA001C. Addition support
 for this work was provided by NASA through an award issued by
 JPL/Caltech. We acknowledge support from NSF CAREER Grant
 AST-0645416. Finally, we would also like to acknowledge the use of
 NASA's High End Computing Program computer systems.

\bibliographystyle{mn2e}
\bibliography{ian}
\label{lastpage}
\end{document}